\documentclass{elsart}

\usepackage{graphicx}
\usepackage{amsmath}
\usepackage{amssymb}

\begin{document}


\begin{frontmatter}

\title{Application of Hamamatsu MPPC to T2K Neutrino Detectors}
\author[KYOTO]{M.~Yokoyama\corauthref{cor}},
\ead{masashi@scphys.kyoto-u.ac.jp}
\author[KYOTO]{T.~Nakaya},
\author[KYOTO]{S.~Gomi},
\author[KYOTO]{A.~Minamino},
\author[KYOTO]{N.~Nagai},
\author[KYOTO]{K.~Nitta},
\author[KYOTO]{D.~Orme},
\author[KYOTO]{M.~Otani},
\author[KEK]{T.~Murakami},
\author[KEK]{T.~Nakadaira},
\author[KEK]{M.~Tanaka},
\author[INR]{Yu.~Kudenko},
\author[TRIUMF]{F.~Retiere},
\author[ICL]{A.~Vacheret}

\address[KYOTO]{Department of Physics, Kyoto University, Kyoto, 606-8502, Japan}
\address[KEK]{IPNS, High Energy Accelerator Research Organization (KEK), Tsukuba, Ibaraki 305-0801, Japan}
\address[INR]{Institute for Nuclear Research, Moscow 117312, Russia}
\address[TRIUMF]{TRIUMF, Vancouver, British Columbia V6T 2A3, Canada}
\address[ICL]{Imperial College, London SW7 2AZ, United Kingdom}

\corauth[cor]{Corresponding author.}

\begin{abstract}
A special type of Hamamatsu MPPC,
with a sensitive area of 1.3$\times$1.3~mm$^2$ containing 667 pixels with 50$\times$50~$\mu$m$^2$ each,
has been developed for the near neutrino detector in the T2K long baseline neutrino experiment.
About 60 000 MPPCs will be used in total to read out the plastic scintillator detectors with wavelength shifting fibers.
We report on the basic performance of MPPCs produced for T2K.
\end{abstract}

\begin{keyword}
Neutrino detector, Giger-mode APD, MPPC
\end{keyword}

\end{frontmatter}

\section{Introduction}
T2K (Tokai-to-Kamioka)~\cite{T2K} is a long baseline neutrino oscillation experiment in Japan, using an intense beam from J-PARC accelerator at Tokai and the massive Super-Kamiokande detector 295~km away.
The main goals of T2K are a sensitive search for the $\nu_e$ appearance from $\nu_\mu$, 
which is related to the mixing angle $\theta_{13}$, and precise measurements of `atmospheric' oscillation parameters.
In order to achieve the aimed precision, good understanding of the beam properties and $\nu$-nucleus interaction are indispensable.
The `near detector' (ND) complex will be placed in Tokai to provide this information.

The T2K-ND~\cite{T2K-ND280} consists of several sub-detectors with specific and complimentary functions.
As the basic elements for particle detection, most of detectors will use the plastic scintillator read out by wavelength shifting (WLS) fibers.
This is a widely used technique, especially in recent accelerator neutrino experiments~\cite{sci-wls}.
In those experiments, multi-anode PMTs (MAPMTs) have been used as the photosensor.
For T2K, MAPMT is not a good candidate because some of detectors have to operate inside a magnetic field of 0.2~T and cope with a limited space available. 

The following are major requirements for photosensors in T2K:
\begin{itemize}
\item More or equal photon detection efficiency than that of a MAPMT.
\item Compact to fit the limited space inside the magnet.
\item Operational in a magnetic field.
\item Good stability and low cost for a large number of readout channels (60 000).
\end{itemize}%
We decided to use the Multi-Pixel Photon Counter (MPPC)~\cite{MPPC, MPPC2} in August 2005. 
Since then, we continued the development collaborating with Hamamatsu and KEK Detector Technology Project.
In this paper, we report on the performance of the MPPC developed for T2K.

\section{MPPC for T2K}
\begin{table}[tbp]
\begin{center}
\caption{Specifications of T2K-MPPC(S10362-13-050C).}
\begin{tabular}{ccc} \hline \hline
\multicolumn{2}{c}{Item} & Spec. \\ \hline
\multicolumn{2}{c}{Active area} & 1.3$\times$1.3~mm$^2$ \\
\multicolumn{2}{c}{Pixel size} & 50$\times$50~$\mu$m$^2$ \\
\multicolumn{2}{c}{Number of pixels} & 667 \\
\multicolumn{2}{c}{Operation voltage} & 70~V (typ.) \\
\multicolumn{2}{c}{PDE @ 550~nm} & $>$15\% \\
Dark count &($>$0.5~pe) & $<$1.35~Mcps\\
$[$ @ 25$^\circ$C $]$  & ($>$1.2~pe) & $<$0.135~Mcps \\
\hline \hline
\end{tabular}
\label{tab:spec}
\end{center}
\end{table}%

\begin{figure}[tbp]
\begin{center}
\includegraphics[width=0.4\textwidth]{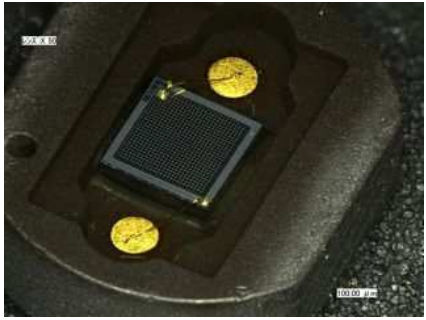}
\caption{MPPC developed for T2K.}
\label{fig:MPPC}
\end{center}
\end{figure}%

The major specifications of the MPPC for the T2K-ND is summarized in Table~\ref{tab:spec}.
Based on the past experience, we use the 1.0~mm diameter Kuraray Y11(200)MS WLS fiber.
We enlarged the sensitive area of the MPPC from 1$\times$1~mm$^2$ of those on catalogue to 1.3$\times$1.3~mm$^2$ so that we can minimize the light loss at the optical contact with a simple coupler.
The size of APD pixel is 50$\times$50~$\mu$m.
In order to make the MPPC fit inside the package, one of the bonding pad needs to be located at the corner of otherwise sensitive area (Fig.~\ref{fig:MPPC}).
The number of APD pixels is 667.
Thanks to the pulsed beam timing, large light yield, and coincidence usable to select a particle trajectory, our requirement on the dark noise rate is not very demanding and about 1~Mcps at 0.5~p.e.\ threshold is acceptable.

\section{Performance of T2K-MPPC}
The delivery of mass production MPPC for T2K was started in February 2008.
By the middle of June 2008, more than 30 000 MPPCs were received by the T2K group.
Among them, about 8000 MPPCs have been measured by the test facility at Kyoto University~\cite{mass-test}.
The measured performance of 5820 MPPCs are presented here.
While we characterize all MPPCs at 15, 20 and 25~$^\circ$C, only results with 25~$^\circ$C are shown here.
More information on the measurement setup and procedure are given in~\cite{mass-test}.

\subsection{Gain and breakdown voltage}

\begin{figure}[tbp]
\begin{center}
\includegraphics[width=0.4\textwidth]{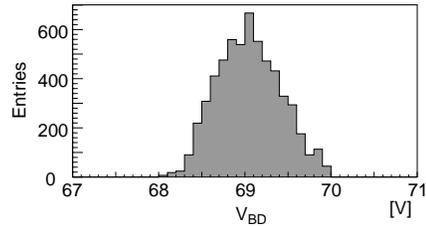}
\caption{Breakdown voltage.}
\label{fig:vbd}
\end{center}
\end{figure}%

The breakdown voltage $V_\mathrm{BD}$, the applied voltage above which the APD operates in the Geiger mode, can be measured from the gain-voltage relation.
The distribution of $V_\mathrm{BD}$ is shown in Fig.~\ref{fig:vbd}.
Note that we selected MPPCs with similar $V_\mathrm{BD}$ here, based on Hamamatsu data sheet, to minimize the necessary range of the bias voltage variation during a measurement.
We plan to use MPPCs with similar $V_\mathrm{BD}$ in one subsystem.
The full width of $V_\mathrm{BD}$ for all MPPCs produced so far is about 3~V, well within the voltage range adjustable by our electronics~\cite{T2K-elec}.

\begin{figure}[tbp]
\begin{center}
\includegraphics[width=0.4\textwidth]{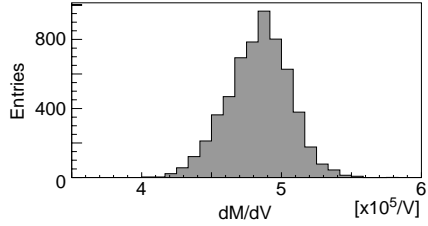}%
\caption{Gain slope.}
\label{fig:cap}
\end{center}
\end{figure}%
Figure~\ref{fig:cap} shows the distribution of measured gain ($M$) slope against the applied voltage (V) for 5820 MPPCs.
The overvoltage $\Delta V$ is defined as the difference between the applied voltage and the breakdown voltage, $\Delta V \equiv V-V_\mathrm{BD}$. 
The typical gain of MPPC is measured to be about 5$\times10^5$ at $\Delta V=1.0V$.
The RMS of the distribution is 4.5\%, showing excellent device uniformity.

\subsection{Dark noise rate}

\begin{figure}[tbp]
\begin{center}
\includegraphics[width=0.4\textwidth]{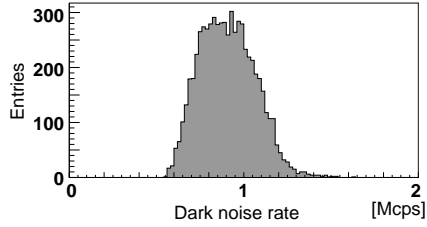}
\caption{Dark noise rate.}
\label{fig:dark}
\end{center}
\end{figure}%

The measured dark noise rate at a threshold of 0.5~p.e.\ is shown in Fig.~\ref{fig:dark}.
It is found that the dark noise rate has relatively large device dependence.
Note that the dark noise rate is used to select MPPCs before shipping from Hamamatsu.

\subsection{Photon detection efficiency}

\begin{figure}[htbp]
\begin{center}
\includegraphics[width=0.4\textwidth]{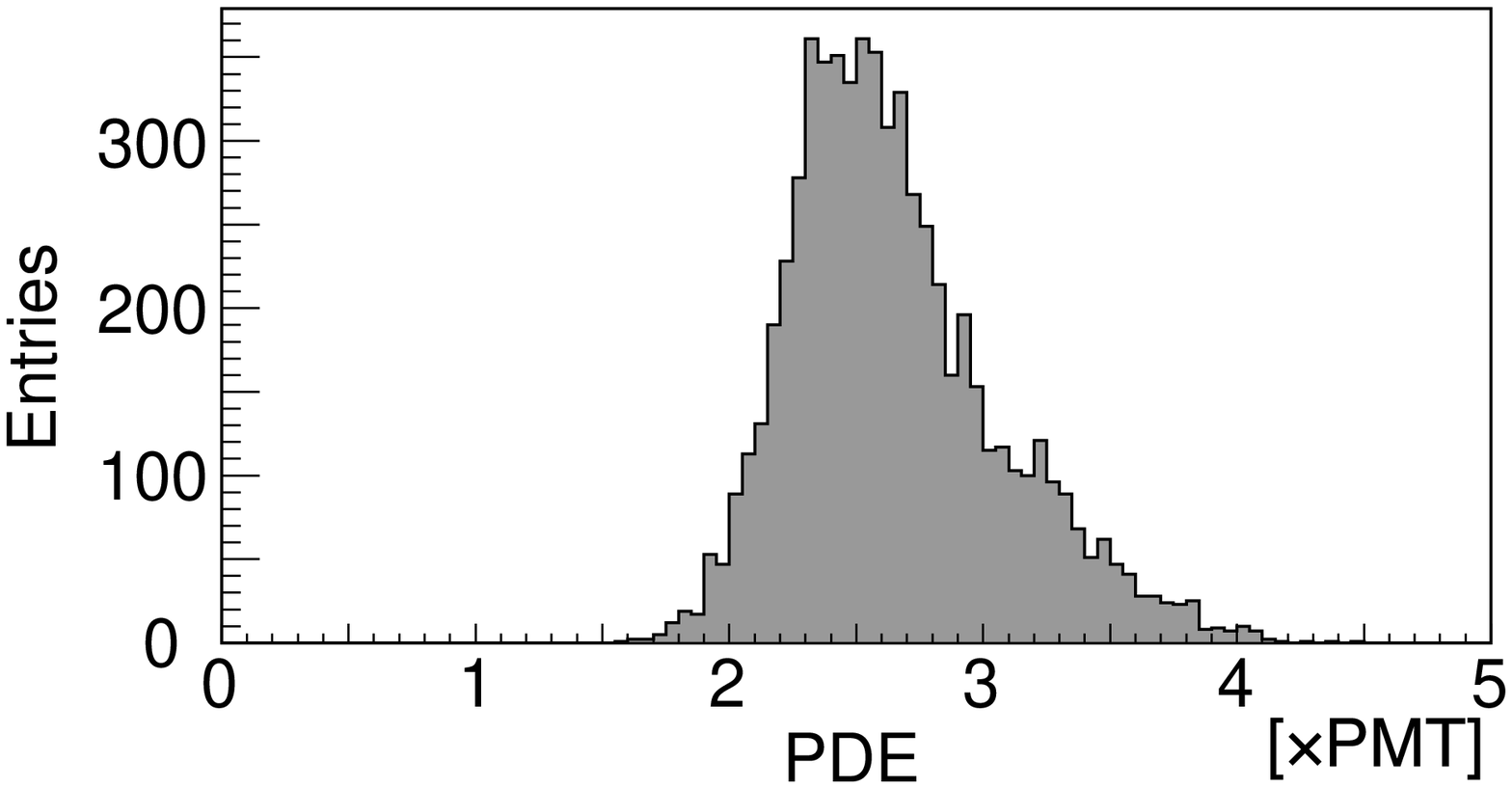}
\caption{PDE.}
\label{fig:pde}
\end{center}
\end{figure}%

The photon detection efficiency (PDE) is measured using a PMT (R1818) as a reference.
A plastic optical coupler~\cite{Gomi} is used for the connection of WLS fiber and MPPC.
The effective PDE, including the effect of the optical coupling, is measured and shown in Fig.~\ref{fig:pde}.
The tail at the larger PDE value is presumably due to the systematics in the light yield measurement for the reference PMT and being checked.
The average PDE is about 2.5 times that of PMT at $\Delta$V=1.5~V.

\subsection{Cross-talk and afterpulse rate}

\begin{figure}[htbp]
\begin{center}
\includegraphics[width=0.4\textwidth]{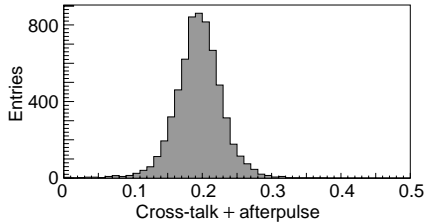}
\caption{Cross-talk and afterpulse rate.}
\label{fig:xtalk}
\end{center}
\end{figure}

The cross-talk and afterpulse rates are measured together from the ADC distribution~\cite{mass-test}.
A detailed study of these effects in MPPC is also given in \cite{Fabrice}.
Figure~\ref{fig:xtalk} shows the measured values for 5820 MPPCs.
The average cross-talk and afterpulse rate is 20\% with $\Delta$V=1.5~V.
The performance is sufficient for our use, although suppression of cross-talk and afterpulse are desired for some other applications.

\subsection{Light yield}
We have measured the light yield with real plastic scintillators to be used in T2K.
Extruded scintillators made at Fermilab, with dimensions 5$\times$120$\times$1~cm$^3$, were exposed to a 3~GeV electron beam at KEK.
A WLS fiber is inserted into the hole at the center of the bar.
No optical grease nor cement was used between the scintillator, fiber, and MPPC.
A plastic coupler~\cite{Gomi} is used to connect MPPC and WLS fiber.
At the center of the bar, the average light yield was about 15~p.e.

The scintillator slabs made in Russia was also measured.
A WLS fiber is embedded in a `S'-shaped groove on the surface of 167$\times$870$\times$7~mm$^3$ scintillator.
Summing signals from both ends of WLS fiber, 36$\pm$3 p.e.\ was obtained for a cosmic ray muon passing through the center of scintillator.

\section{Conclusion}
New type of MPPCs have been developed for the T2K experiment.
The T2K-MPPC is designed to have 1.3$\times$1.3~mm$^2$ sensitive area in order to minimize the light loss when coupled to a WLS fiber of 1.0~mm diameter.
The size of pixel is 50$\times$50~$\mu$m$^2$ and the number of pixels is 667.

In T2K, about 60 000 MPPCs will be used in total.
The mass production has started in February 2008.
The gain, breakdown voltage, noise rate, photon detection efficiency, and cross-talk and afterpulse rate of T2K-MPPCs are measured for each device.
The device uniformity is found to be excellent based on the measurement of 5820 MPPCs.
All of MPPCs satisfy our requirements.
We have established techniques necessary for a large scale application of MPPC to the WLS readout.

The T2K experiment is planed to start in 2009. It will be the first experiment to use MPPCs in a large scale.

\section*{Acknowledgments}
The authors are grateful to the solid state division of Hamamatsu Photonics for providing us test samples during the development.
The development of MPPC is supported by KEK detector technology project.

\end{document}